\documentclass[aps,twocolumn]{revtex4}
\usepackage{amssymb}

\usepackage{graphicx}

\def\b{\begin{equation}}
\def\e{\end{equation}}
\def\balll{\begin{array}{lll}}
\def\ea{\end{array}}
\def\bea{\begin{eqnarray}}
\def\eea{\end{eqnarray}}

\newcommand{\be}{\begin{equation}}
\newcommand{\ee}{\end{equation}}
\newcommand{\bqn}{\begin{eqnarray}}
\newcommand{\eqn}{\end{eqnarray}}

\begin{document}

\title{Addendum: Behavior of a bipartite system in a cavity}
\author{C. A. Linhares$^a$, A. P. C. Malbouisson$^b$ and J. M. C.
Malbouisson$^c$}
\date{\today }

\begin{abstract}
This note is an Addendum to our previous article [Phys. Rev. A
\textbf{81}, 053820 (2010)]. We show that under the assumption of 
a Bose-Einstein distribution for the thermal reservoir, zero-temperature properties
of the entangled states considered there are not changed by heating, for temperatures up to the order of room temperatures.
In this case, the system is dissipative in free space and presents stability for a
small cavity, both for $T=0$ and for finite temperature.
\end{abstract}

\maketitle

\affiliation{\ $^a$Instituto de F\'{\i}sica, Universidade do Estado
do Rio de Janeiro, 20559-900, Rio de Janeiro, RJ, Brazil\\$^b$Centro
Brasileiro de Pesquisas F\'{\i}sicas/MCT, 22290-180, Rio de Janeiro,
RJ, Brazil\\$^c$Instituto de F\'{\i}sica, Universidade Federal da
Bahia, 40210-340, Salvador, BA, Brazil}

%%%%%%%%%%%%%%%%%%%%%%%%%%%%%%%%%%%%%%%%%%%%%%%%%%%%%%%%%%%%%%%%%%%%%
%%%%%%%%%%%%%%%%%%%%%%%%%%%%%%%%%%%%%%%%%%%%%%%%%%%%%%%%%%%%%%%%%%%%%

%%%%%%%%%%%%%%%%%%%%%%%%%%%%%%%%%%%%%%%%%%%%%%%%%%%%%%%%%%%%%%%%%%%%%%%%
%%%%%%%%%%%%%%%%%%%%%%%%%%%%%%%%%%%%%%%%%%%%%%%%%%%%%%%%%%%%%%%%%%%%%%%

%\pacs{...}

%%%%%%%%%%%%%%%%%%%%%%%%%%%%%%%%%%%%%%%%%%%%%%%%%%%%%%%%%%%%%%%%%%%%%
%%%%%%%%%%%%%%%%%%%%%%%%%%%%%%%%%%%%%%%%%%%%%%%%%%%%%%%%%%%%%%%%%%%%%

\textit{Introduction}: In this note we consider how heating could
affect the properties of the bipartite system studied
in~\cite{emaranho}. This bipartite system consists of two dressed
atoms inside a reflecting cavity which do not interact with each
other, only with an environmental field. Our approach to this
problem makes use of the notion of \textit{dressed thermal
states}~\cite{termica}, in the context of a model already employed
in the literature, of atoms, or more generally material particles,
in the harmonic approximation, coupled to an environment modeled by
an infinite set of pointlike harmonic oscillators (the field modes).
The dressed thermal state approach is an extension of the dressed
(zero-temperature) formalism which has been introduced earlier. The
reader is referred to~\cite{emaranho,termica} for details of our
formalism.

We here consider entanglement as a pure quantum effect, a
characteristic of quantum mechanics, which is also nonlocal, in the
sense that distant and noninteracting systems may be entangled. This
is due to the physical meaning attributed to superposed states, a
concept with no correspondence in classical physics, and not to the
interaction between the (in our case, dressed) atoms. Indeed, such
properties of entanglement of noninteracting systems have been used
in the realm of teleportation and quantum information theory and to
conceive quantum communication devices. We think that the
investigation of in which measure the properties of such systems are
affected by heating is important, particularly in the case of room
temperatures. The possibility of constructing such devices at
temperatures of everyday life would be an interesting matter.

In this note we remain on theoretical grounds, and perform a study
of  thermal effects on our bipartite system, assuming that the
dressing fields of the atoms obey the Bose-Einstein distribution and
that the first-level excited dressed atomic states evolve in the
same way as in~\cite{emaranho}. Under these assumptions, it results
that the reduced density matrix does not depend on temperature. On
the other hand, employing methods and relying on results obtained
in~\cite{termica,termalizacao}, one finds that the individual
dressed atoms in the bipartite system are very little affected by
heating for temperatures around $300\,\rm{K}$. The overall
conclusion is that, methods and results from~\cite{emaranho} are
valid if we consider the system at room temperatures.

Our bipartite system is composed of two subsystems, $\mathcal{A}$
and $\mathcal{B}$; the subsystems consist respectively of dressed
atoms $A$ and $B $ and the whole system is contained in a perfectly
reflecting sphere of radius $R$ in thermal equilibrium with an
environment (field) at a temperature $\beta^{-1}$. We consider each
atom carrying its own dressing field cloud, independently of each
other. This means that we are considering the behavior of a
noninteracting bipartite system. Also, we take an specific type of
reservoir, in the thermal Bose-Einstein state, where the excitation
distribution is diagonal in the number basis.

\textit{The density matrix}: We start, at time $t=0$, from a family of
superposed states of the bipartite system given by
\begin{widetext}
\begin{equation}
\left| \Psi _{\mathcal{AB}}\right\rangle =\sqrt{\xi }\,\left| \Gamma _{10}
 ^{(AB)}(0)\right\rangle
+\sqrt{1-\xi }\,e^{i\phi }\,\left|
\Gamma _{01}^{(AB)}(0)\right\rangle
%\nonumber \\
\equiv \sqrt{\xi }\,\left| 1_A0_B\right\rangle
+\sqrt{
1-\xi }\,e^{i\phi }\,\left| 0_A1_B\right\rangle ,
 \label{definition-entangled-state}
\end{equation}
\end{widetext}
where $0\leq \xi \leq 1$.

In Eq.~(\ref{definition-entangled-state}), $\left| \Gamma
_{10}^{(AB)}(0)\right\rangle \equiv \left| 1_A0_B\right\rangle $ and $\left|
\Gamma _{01}^{(AB)}(0)\right\rangle \equiv \left| 0_A1_B\right\rangle $
stand respectively for the $t=0$ states in which the dressed atom $A$ ($B$)
is at the first level, the dressed atom $B$ ($A$) in the ground state. They
are
\begin{equation}
\left| \Gamma _{10}^{(AB)}(0)\right\rangle =\left| \Gamma
_1^A(0)\right\rangle \otimes \left| \Gamma _0^B(0)\right\rangle
\end{equation}
and
\begin{equation}
\left| \Gamma _{01}^{(AB)}(0)\right\rangle =\left| \Gamma
_0^A(0)\right\rangle \otimes \left| \Gamma _1^B(0)\right\rangle .
\end{equation}

In order to take into account temperature effects, we consider the density
matrix,
\[
\rho _{\mathcal{AB}}(t,\beta )=\left| \Psi _{\mathcal{AB}}\right\rangle
\left\langle \Psi _{\mathcal{AB}}\right| \otimes \rho _{\mathcal{A}}(\beta
)\otimes \rho _{\mathcal{B}}(\beta ),
\]
in terms of dressed objects. We take the density matrix for the
thermal bath obeying the Bose-Einstein distribution,
\begin{eqnarray}
\rho _{\mathcal{A}}(\beta ) &=&\prod_k\frac 1{Z_k}e^{-\hbar \beta \omega
_ka_k^{\dagger }a_k}  \nonumber \\
&=&\prod_k\frac 1{Z_k}\sum_{n_k=0}^\infty e^{-\hbar \beta \omega
_kn_k}\left| n_k\right\rangle \left\langle n_k\right| ,
\end{eqnarray}
where $n_k=0,1,2,\ldots $ correspond to the occupation numbers of the
dressed field modes $k$, and $Z_k$ is obtained imposing the condition of
unit trace for $\rho _{\mathcal{A}}(\beta )$; this gives
\begin{equation}
Z_k=\frac 1{1-e^{-\hbar \beta \omega _kn_k}},
\end{equation}
and so,
\begin{widetext}
\begin{eqnarray}
\rho _{\mathcal{A}}(\beta ) & = &\left( \frac
1{Z_1}\sum_{n_1}e^{-\hbar \beta \omega _1n_1}\left| n_1\right\rangle
\left\langle n_1\right| \right) \otimes \left( \frac
1{Z_2}\sum_{n_2}e^{-\hbar \beta \omega _2n_2}\left|
n_2\right\rangle \left\langle n_2\right| \right) \otimes \cdots \nonumber \\
& = &\sum_{n_1,n_2,\ldots }\frac{e^{-\hbar \beta (\omega
_1n_1+\omega _2n_2+\cdots )}}{Z_1Z_2\cdots }\left| n_1\right\rangle
\left\langle n_1\right| \otimes \left| n_2\right\rangle \left\langle
n_2\right| \otimes \cdots . \label{rhobeta}
\end{eqnarray}

For the part containing only the atoms,
we have
\begin{eqnarray}
\left| \Psi _{\mathcal{AB}}\right\rangle \left\langle \Psi
_{\mathcal{AB} }\right|  &=&\xi \left[ \left| 1_A(t)\right\rangle
\left\langle 1_A(t)\right| \otimes \left| 0_B\right\rangle
\left\langle 0_B\right|
\right] \nonumber \\
&&+(1-\xi )\left[ \left| 0_A\right\rangle \left\langle 0_A\right| \otimes
\left| 1_B(t)\right\rangle \left\langle 1_B(t)\right| \right] \nonumber \\
&&+\sqrt{\xi (1-\xi )}e^{i\phi }\left[ \left| 0_A\right\rangle \left\langle
1_A(t)\right| \otimes \left| 1_B(t)\right\rangle \left\langle 0_B\right|
\right] \nonumber \\
&&+\sqrt{\xi (1-\xi )}e^{-i\phi }\left[ \left| 1_A(t)\right\rangle
\left\langle 0_A\right| \otimes \left| 0_B\right\rangle \left\langle
1_B(t)\right| \right],
\label{rhoatoms}
\end{eqnarray}
\end{widetext}
where we have changed notation, $\left| \Gamma _1^A(t)\right\rangle \equiv
\left| 1_A(t)\right\rangle $, and similarly for the other states.

Putting together Eqs.~(\ref{rhobeta}) and (\ref{rhoatoms}), the
reduced density matrix is obtained by taking the trace over the
field modes,
\begin{widetext}
\begin{eqnarray}
\rho _{p_Ap_B}^{r_Ar_B}(t,\beta )
&=&\xi \sum_{\{n_i\}}\frac{e^{-\hbar \beta \sum \omega _in_i}}{\prod Z_i}
\sum_{\left\{ k_i\right\} =1}^\infty \left\langle p_A,k_1,\ldots
|1_A(t);n_1,\ldots \right\rangle \left\langle 1_A(t);n_1,\ldots
|r_A,k_1,\ldots \right\rangle \nonumber \\
&& \times \sum_{\{m_i\}}\frac{e^{-\hbar \beta \sum \omega
_im_i}}{\prod Z_i} \sum_{\left\{ q_i\right\} =1}^\infty \left\langle
p_B,q_1,\ldots |0_B;m_1,\ldots \right\rangle \left\langle
0_B;m_1,\ldots |r_B,q_1,\ldots
\right\rangle \nonumber \\
&&+\left( 1-\xi \right) \sum_{\{n_i\}}\frac{e^{-\hbar \beta \sum \omega
_in_i}}{\prod Z_i}\sum_{\left\{ k_i=1\right\} }^\infty \left\langle
p_A,k_1,\ldots |0_A;n_1,\ldots \right\rangle \left\langle 0_A;n_1,\ldots
|r_A,k_1,\ldots \right\rangle \nonumber \\
&& \times \sum_{\{m_i\}}\frac{e^{-\hbar \beta \sum \omega
_im_i}}{\prod Z_i} \sum_{\left\{ q_i\right\} =1}^\infty \left\langle
p_B,q_1,\ldots |1_B(t);m_1,\ldots \right\rangle \left\langle
1_B(t);m_1,\ldots
|r_B,q_1,\ldots \right\rangle \nonumber \\
&&+\sqrt{\xi (1-\xi )}e^{i\phi }\sum_{\{n_i\}}\frac{e^{-\hbar \beta \sum
\omega _in_i}}{\prod Z_i}\sum_{\left\{ k_i\right\} =1}^\infty \left\langle
p_A,k_1,\ldots |0_A;n_1,\ldots \right\rangle \left\langle 1_A(t);n_1,\ldots
|r_A,k_1,\ldots \right\rangle \nonumber \\
&& \times \sum_{\{m_i\}}\frac{e^{-\hbar \beta \sum \omega
_im_i}}{\prod Z_i} \sum_{\left\{ q_i\right\} =1}^\infty \left\langle
p_B,q_1,\ldots |1_B(t);m_1,\ldots \right\rangle \left\langle
0_B;m_1,\ldots |r_B,q_1,\ldots
\right\rangle \nonumber \\
&&+\sqrt{\xi (1-\xi )}e^{-i\phi }\sum_{\{n_i\}}\frac{e^{-\hbar \beta \sum
\omega _in_i}}{\prod Z_i}\sum_{\left\{ k_i\right\} =1}^\infty \left\langle
p_A,k_1,\ldots |1_A(t);n_1,\ldots \right\rangle \left\langle 0_A;n_1,\ldots
|r_A,k_1,\ldots \right\rangle \nonumber \\
&& \times \sum_{\{m_i\}}\frac{e^{-\hbar \beta \sum \omega
_im_i}}{\prod Z_i} \sum_{\left\{ q_i\right\} =1}^\infty \left\langle
p_B,q_1,\ldots |0_B;m_1,\ldots \right\rangle \left\langle
1_B(t);m_1,\ldots |r_B,q_1,\ldots \right\rangle . \label{reduzida}
\end{eqnarray}
\end{widetext}

To calculate the matrix elements above we use the time evolution of the
states involved~\cite{emaranho},
\begin{eqnarray}
\left| 1_A(t);n_1,n_2,\ldots \right\rangle =\sum_\nu f_{A\nu }(t)\left|
1_\nu (0);n_1,n_2,\ldots \right\rangle ,  \label{evolucao}
\end{eqnarray}
where the coefficients $f_{A\nu }(t)$ are
\begin{equation}
f_{A\nu }(t)=\sum_st_\mu ^st_\nu ^se^{-i\Omega _st};\;\;\;\sum_\nu \left|
f_{A\nu }(t)\right| ^2=1\,.  \label{ortovestidas5}
\end{equation}
A similar equation holds for $\left| 1_B(t);n_1,n_2,\ldots
\right\rangle $. Inserting Eq.~(\ref{evolucao}) and its equivalent
for the atom $B$ into Eq.~\ref{reduzida}), we obtain, after some
rather long manipulations,
\begin{widetext}
\begin{eqnarray*}
\rho _{p_Ap_B}^{r_Ar_B}(t) &=&\left( \sum_{\{n_i\}}\frac{e^{-\hbar \beta
\sum \omega _in_i}}{\prod Z_i}\right) \left( \sum_{\{m_i\}}\frac{e^{-\hbar
\beta \sum \omega _im_i}}{\prod Z_i}\right)
%\\
%&&\times
\left\{ \xi \left[ \left| f_{AA}(t)\right| ^2\delta _{p_A1}\delta
_{r_A1}+\sum_{i=1}^\infty \left| f_{Ai}(t)\right| ^2\delta _{p_A0}\delta
_{r_A0}\right] \delta _{p_B0}\delta _{r_B0}\right. \nonumber \\
&&+\left( 1-\xi \right) \delta _{p_A0}\delta _{r_A0}\left[ \left|
f_{BB}(t)\right| ^2\delta _{p_B1}\delta _{r_B1}+\sum_{i=1}^\infty \left|
f_{Bi}(t)\right| ^2\delta _{p_B0}\delta _{r_B0}\right]
\nonumber \\
&&
+\sqrt{\xi \left( 1-\xi \right) }e^{i\phi }f_{AA}^{*}(t)f_{BB}(t)\delta
_{p_A0}\delta _{r_A1}\delta _{p_B1}\delta _{r_B0}
%\nonumber \\
%&&
+\left. \sqrt{\xi \left( 1-\xi \right) }e^{-i\phi
}f_{AA}(t)f_{BB}^{*}(t)\delta _{p_A1}\delta _{r_A0}\delta _{p_B0}\delta
_{r_B1}\right\} .
\label{matrizfinal}
\end{eqnarray*}
\end{widetext}
But
\begin{equation}
\frac 1{Z_i}\sum_{n_i}e^{-\hbar \beta \omega _in_i}=1,\qquad \forall i,
\label{norm}
\end{equation}
and, therefore, the temperature-dependent terms disappear, leaving
the reduced density matrix with exactly the same form as
in~\cite{emaranho}. For identical atoms we have
$f_{AA}(t)=f_{BB}(t)\equiv f_{00}(t)$ and the nonvanishing matrix
elements of $\rho $ are
\begin{eqnarray}
\rho _{00}^{00}(t) &=&1-\left| f_{00}(t)\right| ^2,  \nonumber \\
\rho _{01}^{01}(t) &=&(1-\xi )\left| f_{00}(t)\right| ^2,  \nonumber \\
\rho _{10}^{10}(t) &=&\xi \left| f_{00}(t)\right| ^2,  \label{elementos} \\
\rho _{01}^{10}(t) &=&\sqrt{\xi (1-\xi )}e^{i\phi }\left| f_{00}(t)\right|
^2,  \nonumber \\
\rho _{10}^{01}(t) &=&\sqrt{\xi (1-\xi )}e^{-i\phi }\left| f_{00}(t)\right|
^2.  \nonumber
\end{eqnarray}

Notice that this property is not specific to the Bose-Einstein
distribution we have employed, but it is common to any normalized
distribution which is diagonal in the number basis. This may not
happen for other types of reservoir, like coherent squeezed thermal
states, where the factorization of the temperature-dependent terms
may not occur.

\textit{Conclusions}: At finite temperature, all the results that
depend only on the reduced density matrix remain the same as in the
zero-temperature case~\cite{emaranho}. This conclusion is valid for
our choice of the Bose-Einstein distribution to the heated
environment, using Eq.~(\ref{evolucao}) to the time evolution of the
first-level excited atomic states. The choice of the Bose-Einstein
distribution to the field modes can be justified: in the case of an
arbitrarily large cavity, the dressed field modes coincide with the
bare ones~\cite {termica} and in the limit of vanishing coupling
with the atom, these modes follow the Bose-Einstein distribution
exactly. Strictly speaking, this is not the case for the coupled
atom-field system in a finite cavity. Nevertheless, in many
situations this approximation is acceptable in the weak-coupling
regime~\cite{footnote}. The use of Eq.~(\ref{evolucao}) to the time
evolution of the first-level excited atomic states, as we will see
below, is a good approximation for room temperatures.

In this case, heating does affect each atom individually.
In~\cite{termalizacao} and in~\cite{termica}, some of us have
studied the time evolution of the temperature-dependent occupation
number of a single dressed atom, $n_0^{\prime }(t,\beta )$, for
respectively a very large cavity (free space) and confined in a
small cavity, with the same heating procedure as we have used here.
In the case of a large cavity, we have shown that, starting from an
initial value $n_0^{\prime }(t=0,\beta )=1$, $n_0^{\prime }(t,\beta
) $ evolves, for weak coupling, steadily from $1$ to an equilibrium
value of
\[
n_0^{\prime }(\infty ,\beta )=1/(e^{\hbar \beta \bar{\omega}}-1).
\]
For $T=300\,\mathrm{K}$, for a frequency $\bar{\omega}=4.0\times
10^{14}$/s, we have $n_0^{\prime }(\infty ,T=300)\approx 0.09$, a
value slightly higher than the corresponding value for $T=0$,
$n_0^{\prime }(\infty ,T=0)=0$. For the same initial condition, $n_0^{\prime }(t=0,\beta )=1$, and the same value of the emission
frequency and temperature as in the large cavity, it is found
in~\cite{termica} for a small cavity (radius $R\sim 10^{-6}${\ m})
that the time evolution of the occupation number has an oscillating
behavior. Raising the temperature increases the average value of the
occupation number, but this increase is significative only for high
laboratory temperatures. For instance, for $T=10^5$K ($\sim 8.4$ eV)
(a value lower than the ionization temperature of $13.6$ eV for the
hydrogen atom) for a frequency $\bar{\omega}=4.0\times 10^{14}$/s,
as above, one obtains that the average occupation number is about
$3.5$ times higher than the zero-temperature value, $n_0^{\prime
}(t,T=0)\approx 1$. On the other hand, this effect is negligible for
room temperatures; in this case, taking for instance a temperature
$T\gtrsim 300\,\mathrm{K}$, it is found that the occupation number
is given by $n_0^{\prime }(t,T\gtrsim 300)\gtrsim 1$, which will
remain very close to the zero-temperature value. Therefore, if we
consider the system at room temperatures, individual atoms are very
little affected by heating and the use of Eq.(\ref{evolucao}) is justified. 

Our system is composed of two noninteracting atoms, each one
carrying its own dressing field.  The superposition principle
dictates the existence of correlations between the parts of the
system (the two atoms) even if they do not interact.
By introducing temperature in the environment by means of the
Bose-Einstein distribution, the result is that the entanglement
properties of the considered bipartite system are not affected by
heating up to room temperatures. In particular, considering two identical atoms, the
concurrence, the entanglement of formation and the negativity depend
only on the matrix elements in Eq.~(\ref{elementos}) and thus,
cannot depend on the temperature. However, the temperature of the
environment cannot be so high that the atoms dissociate, or such
that the adopted atomic model becomes invalid. More precisely, from
a physical point of view, our model applies for room temperatures.
As we have seen above, for temperatures of the order of $\sim 300$ K
or not much higher, the harmonic approximation we have used
in~\cite{emaranho} remains valid and the time evolution of the thermal occupation number
of the atoms is very close to the zero-temperature case. This
establishes the range of validity of the present study.

\end{document}